\documentclass[letterpaper]{article} % DO NOT CHANGE THIS
\usepackage{amsmath,amsfonts,bm}

\def\eqref#1{equation~\ref{#1}}
\def\1{\bm{1}}

\def\vs{{\bm{s}}}

\def\vx{{\bm{x}}}
\def\vy{{\bm{y}}}

\def\mE{{\bm{E}}}

\def\mO{{\bm{O}}}

\def\mW{{\bm{W}}}
\def\mX{{\bm{X}}}

\DeclareMathAlphabet{\mathsfit}{\encodingdefault}{\sfdefault}{m}{sl}
\SetMathAlphabet{\mathsfit}{bold}{\encodingdefault}{\sfdefault}{bx}{n}

\usepackage{aaai24}  % DO NOT CHANGE THIS
\usepackage{times}  % DO NOT CHANGE THIS
\usepackage{helvet}  % DO NOT CHANGE THIS
\usepackage{courier}  % DO NOT CHANGE THIS
\usepackage{multirow}
\usepackage[hyphens]{url}  % DO NOT CHANGE THIS
\usepackage{graphicx} % DO NOT CHANGE THIS
\usepackage{booktabs}
\usepackage{tabularx}
\usepackage{amsmath}
\usepackage{csquotes}
\usepackage{amssymb}% http://ctan.org/pkg/amssymb
\usepackage{pifont}% http://ctan.org/pkg/pifont
\usepackage{natbib}  % DO NOT CHANGE THIS AND DO NOT ADD ANY OPTIONS TO IT
\usepackage[scientific-notation=true]{siunitx}
\usepackage{caption} % DO NOT CHANGE THIS AND DO NOT ADD ANY OPTIONS TO IT
\usepackage{algorithm}
\usepackage{algorithmic}
\usepackage{soul} % Required for highlighting

\usepackage{newfloat}
\usepackage{listings}
\DeclareCaptionStyle{ruled}{labelfont=normalfont,labelsep=colon,strut=off} % DO NOT CHANGE THIS
\floatstyle{ruled}
\newfloat{listing}{tb}{lst}{}
\floatname{listing}{Listing}
\title{Learning Temporal Resolution in Spectrogram for Audio Classification}
\author {
    Haohe Liu\textsuperscript{\rm 1},
    Xubo Liu\textsuperscript{\rm 1},
    Qiuqiang Kong\textsuperscript{\rm 2},
    Wenwu Wang\textsuperscript{\rm 1},
    Mark D. Plumbley\textsuperscript{\rm 1}
}
\affiliations{
    \textsuperscript{\rm 1}University of Surrey\\
    \textsuperscript{\rm 2}The Chinese University of Hong Kong\\
}

\usepackage{bibentry}
\begin{document}

\maketitle

\begin{abstract}
The audio spectrogram is a time-frequency representation that has been widely used for audio classification. One of the key attributes of the audio spectrogram is the temporal resolution, which depends on the hop size used in the Short-Time Fourier Transform (STFT). Previous works generally assume the hop size should be a constant value (e.g., 10 ms). However, a fixed temporal resolution is not always optimal for different types of sound. The temporal resolution affects not only classification accuracy but also computational cost. This paper proposes a novel method, \textbf{DiffRes}, that enables \textbf{diff}erentiable temporal \textbf{res}olution modeling for audio classification. Given a spectrogram calculated with a fixed hop size, DiffRes merges non-essential time frames while preserving important frames. DiffRes acts as a ``drop-in'' module between an audio spectrogram and a classifier and can be jointly optimized with the classification task. We evaluate DiffRes on five audio classification tasks, using mel-spectrograms as the acoustic features, followed by off-the-shelf classifier backbones. Compared with previous methods using the fixed temporal resolution, the DiffRes-based method can achieve the equivalent or better classification accuracy with at least 25\% computational cost reduction. We further show that DiffRes can improve classification accuracy by increasing the temporal resolution of input acoustic features, without adding to the computational cost.
\end{abstract}

\section{Introduction}

Audio classification refers to a series of tasks that assign labels to an audio clip. Those tasks include audio tagging~\citep{kong2020panns}, speech keyword classfication~\citep{kim2021broadcasted}, and music genres classification~\citep{castellon2021codified}. The input to an audio classification system is usually a one-dimensional audio waveform, which can be represented by discrete samples. 
Although there are methods using time-domain samples as features~\citep{kong2020panns, lee2017raw}, the majority of studies on audio classification convert the waveform into a spectrogram as the input feature~\citep{gong2021psla, gong2021ast}. 
% Time-frequency features such as spectrogram are widely used for audio classification~\citep{gong2021psla, gong2021ast}.
Spectrogram is usually calculated by the Fourier transform~\citep{champeney1987handbook}, which is applied in short waveform chunks multiplied by a windowing function, resulting in a two-dimensional time-frequency representation. According to the Gabor’s uncertainty principle~\citep{gabor1946theory}, there is always a trade-off between time and frequency resolutions. 
% For example, a longer window size will result in higher frequency resolution but lower temporal resolution. 
To achieve the desired resolution on the temporal dimension, it is a common practice~\citep{kong2021decoupling, liu2022segment} to apply a fixed hop size between windows to capture the dynamics between adjacent frames. With the fixed hop size, the spectrogram has a fixed temporal resolution, which we will refer to simply as resolution in this work. 

\begin{figure}[t!]
    \centering
    \includegraphics[page=1,width=1.0\linewidth]{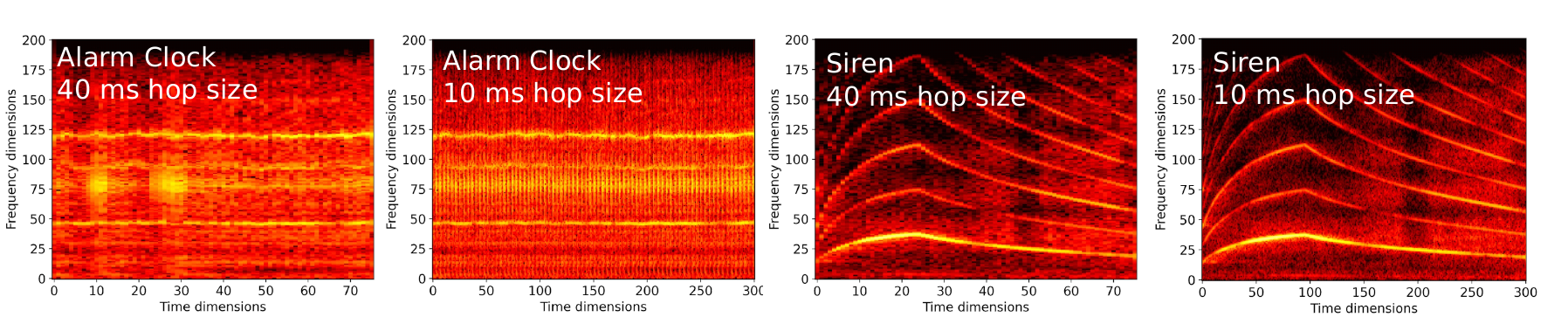}
    \caption{The spectrogram of \textit{Alarm Clock} and~\textit{Siren} sound with $40$~ms and $10$~ms hop sizes. All with a $25$~ms window size. The pattern of \textit{Siren}, which is relatively stable, does not change significantly using a smaller hop size (i.e., larger temporal resolution), while \textit{Alarm Clock} is the opposite.}
    \label{fig:hop-size-compare}
\end{figure}

Using a fixed resolution is not necessarily optimal for an audio classification model. Intuitively, the resolution should depend on the temporal pattern: fast-changing signals are supposed to have high resolution, while relatively steady signals or blank signals may not need the same high resolution for the best accuracy~\citep{huzaifah2017comparison}. For example, Figure~\ref{fig:hop-size-compare} shows that by increasing resolution, more details appear in the spectrogram of \textit{Alarm Clock} while the pattern of \textit{Siren} stays mostly the same. This indicates the finer details in high-resolution \textit{Siren} may not essentially contribute to the classification accuracy. There are plenty of studies on learning a suitable frequency resolution with a similar spirit~\citep{stevens1937scale, sainath2013learning, ravanelli2018speaker, zeghidour2021leaf}. 
Most previous studies focus on investigating the effect of different temporal resolutions~\citep{kekre2012speaker,huzaifah2017comparison,ilyashenko2019trainable,simpf}. 
~\citet{huzaifah2017comparison} observe the optimal temporal resolution for audio classification is class dependent.
~\citet{ferraro2021low} experiment on music tagging with coarse-resolution spectrograms, and observes a similar performance can be maintained while being much faster to compute. 
~\citet{kazakos2021slow} propose a two-stream architecture that processes both fine-grained and coarse-resolution spectrogram and shows the state-of-the-art result on VGG-Sound~\citep{chen2020vggsound}. Recently, \citet{simpf} proposed a non-parametric spectrogram-pooling-based module that can improve classification efficiency with negligible performance degradation. However, these approaches are generally built on a fixed temporal resolution, which is not always optimal for diverse sounds in the world. Intuitively, it is natural to ask: can we dynamically learn the temporal resolution for audio classification?

In this work, we demonstrate the first attempt to learn temporal resolution in the spectrogram for audio classification. We show that learning temporal resolution leads to efficiency and accuracy improvements over the fixed-resolution spectrogram. We propose a lightweight algorithm, DiffRes, that makes spectrogram resolution differentiable during model optimization.  DiffRes can be used as a ``drop-in'' module after spectrogram calculation and optimized jointly with the downstream task.  For the optimization of DiffRes, we propose a loss function, guide loss, to inform the model of the low importance of empty frames formed by SpecAug~\citep{park2019specaugment}. The output of DiffRes is a time-frequency representation with varying resolution, which is achieved by adaptively merging the time steps of a fixed-resolution spectrogram. The adaptive temporal resolution alleviates the spectrogram temporal redundancy and can speed up computation during training and inference.
We perform experiments on five different audio tasks, including the largest audio dataset AudioSet~\citep{gemmeke2017audio}.  DiffRes shows clear improvements on all tasks over the fixed-resolution mel-spectrogram baseline and other learnable front-ends~\citep{zeghidour2021leaf, ravanelli2018speaker, zeghidour2017learning}. 
Compared with methods using fixed-resolution spectrogram, we show that using DiffRes-based models can achieve a computational cost reduction of at least $25$\% with the equivalent or better audio classification accuracy. 

Besides, the potential of the high-resolution spectrogram, e.g., with a one-millisecond hop size, is still unclear. Some popular choices of hop size including $10$~ms~\citep{bock2012online, kong2020panns, gong2021ast} and $12.5$~ms~\citep{rybakov2022real_12_5}. Previous studies~\cite {kong2020panns,ferraro2021low} show classification performance can be steadily improved with the increase of resolution. One remaining question is: can even finer resolution improve the performance?  We conduct an ablation study for this question on a limited-vocabulary speech recognition task with hop sizes smaller than $10$~ms. We noticed that accuracy can still be improved with smaller hop size, at a cost of increased computational complexity. By introducing DiffRes with high-resolution spectrograms, we observe that the classifier performance gains are maintained while the computational cost is significantly reduced.

Our contributions are summarized as follows:
\begin{itemize}
    \item We present DiffRes, a differentiable approach for learning temporal resolution in the audio spectrogram, which improves classification accuracy and reduces the computational cost for off-the-shelf audio classification models.
    \item We extensively evaluate the effectiveness of DiffRes on five audio classification tasks. We further show that DiffRes can improve classification accuracy by increasing the temporal resolution of input acoustic features, without adding to the computational cost.
    \item Our code is available at \url{https://github.com/haoheliu/diffres-python}.
\end{itemize}

\section{Method}

% Every output frame is a weighted average of the input frame
% Given an audio signal$s\in \mathbb{R}^{L}$, the audio classification system $G$ assigns one or multiple labels to $s$. $G$ will first perform feature extraction on $s$ using a function $f$. In the mel-spectrogram based system, feature extraction includes calculating the magnitude spectrogram and multiplying with a mel filterbank: $g_{feat}=W_{N^{\prime}\times N}|\text{STFT}(s; \text{h}, \text{w})|_{N\times M}$, where \text{$STFT$} stands for short-time-fourier-transform, and $W$ is the mel filterbank 
We provide an overview of DiffRes-based audio classification in Section~\ref{sec-overview}. 
% We will use logmel-spectrogram and mel-spectrogram interchangeably. 
We introduce the detailed formulation and the optimization of DiffRes in Section~\ref{sec:Monotonic-Frame-warping}, and~\ref{sec:end-to-end-optimization}. 
% We discuss the regularization and data augmentation effect of DiffRes in Section~\ref{sec:regularization}.

\subsection{Overview}
\label{sec-overview}

\begin{figure*}[t]
    \centering
  \includegraphics[page=1,width=1.0\linewidth]{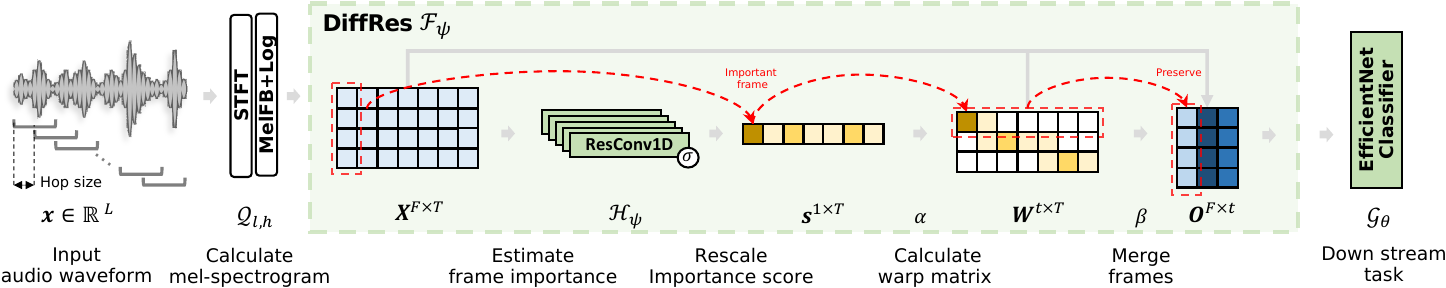}
    \caption{Audio classification with DiffRes and mel-spectrogram. Green blocks contain learnable parameters. DiffRes is a ``drop-in'' module between spectrogram calculation and the downstream task. }
    \label{fig-overview}
\end{figure*}

Let $\vx\in \mathbb{R}^{L}$ denote a one-dimensional audio time waveform, where $L$ is the number of audio samples. An audio classification system can be decomposed into a feature extraction stage and a classification stage. In the feature extraction stage, the audio waveform will be processed by a function $\mathcal{Q}_{l,h}: \mathbb{R}^{L} \rightarrow \mathbb{R}^{F\times T}$,
% \begin{equation}
%     \label{eq:mel-spec}
%     \mathcal{S}(l,h): \mathbb{R}^{L} \rightarrow \mathbb{R}^{f\times t},
% \end{equation}
which maps the time waveform into a two-dimensional time-frequency representation $\mX$, such as a mel-spectrogram, where $\mX_{:,\tau}=(\mX_{1,\tau},...,\mX_{F,\tau})$ is the $\tau\text{-th}$ frame. 
Here, $T$ and $F$ stand for the time and frequency dimensions of the extracted representation. We also refer to the representation along the temporal dimensions as frames. We use $l$ and $h$ to denote window length and hop size, respectively. 
Usually $T\propto\frac{L}{h}$. 
We define the temporal resolution $\frac{1}{h}$ by frame per second~(FPS), which denotes the number of frames in one second. In the classification stage, $\mX$ will be processed by a classification model $\mathcal{G}_{\theta}$ parameterized by $\theta$. 
The output of $\mathcal{G}_{\theta}$ is the label predictions $\hat{\vy}$, in which $\hat{\vy_{i}}$ denotes the probability of class $i$. Given the paired training data $(\vx,\vy)\in \mathbb{D}$, where $\vy$ denotes the one-hot vector for ground-truth labels, the optimization of the classification system can be formulated as
\begin{equation}
    \label{eq:overview}
    \underset{\theta}{\operatorname{arg~min}}~\mathbf{E}_{(\vx,\vy)\sim \mathbb{D}}~{\mathcal{L}(\mathcal{G}_{\theta}(\mX), \vy)},
\end{equation}
where $\mathcal{L}$ is a loss function such as cross entropy~\citep{de2005tutorial}. Figure~\ref{fig-overview} 
% and Algorithm~\ref{algorithm:learnable_pooling} 
show an overview of performing classification with DiffRes. DiffRes is a ``drop-in'' module between $\mX$ and $\mathcal{G}_{\theta}$ focusing on learning the optimal temporal resolution with a learnable function $\mathcal{F}_{\phi}: \mathbb{R}^{F\times T} \rightarrow \mathbb{R}^{F \times t}$, 
% \begin{equation}
%     \label{eq:feature-learning}
%     % ~\mathcal{F}_{\phi}=\alpha \circ \beta
% \end{equation} 
where $t$ is the parameter denoting the target output time dimensions of DiffRes, and $\phi$ is the learnable parameters. DiffRes formulates $\mathcal{F}_{\phi}$ with two steps: 
i) estimating the importance of each time frame with a learnable model $\mathcal{H}_{\phi}$: $\mX \rightarrow \vs$, where $\vs$ is a $1\times T$ shape row vector; and ii) warping frames based on a frame warping algorithm, the warping is performed along a single direction on the temporal dimension. We introduce the details of these two steps in Section~\ref{sec:frame-importance-estimation}. We define the \textit{dimension reduction rate}~$\delta$ of DiffRes by $\delta=(T-t)/T$. 
Usually, $\delta \leq 1$ and $t \leq T$ because the temporal resolution of the DiffRes output is either coarser or equal to that of $\mX$. Given the same $T$, a larger $\delta$ means fewer temporal dimensions $t$ in the output of DiffRes, and usually less computation is needed for $\mathcal{G}_{\theta}$. Similar to Equation~\ref{eq:overview}, $\mathcal{F}_{\phi}$ can be jointly optimized with $\mathcal{G}_{\theta}$ by
\begin{equation}
    \label{eq:overview-with-DiffRes}
    \underset{\theta,\phi}{\operatorname{arg~min}}~\mathbf{E}_{(\vx,\vy)\sim \mathbb{D}}~{\mathcal{L}(\mathcal{G}_{\theta}(\mathcal{F}_{\phi}(\mX)), \vy)}.
\end{equation}
% $\mathcal{F}_{\phi}$ in Equation~\ref{eq:feature-learning} is end-to-end optimized with Equation~\ref{eq:overview} to achieve the optimal performance with fewer time dimensions.

% If we use mel-spectrogram as the input feature of $G$, we will first compute the magnitude spectrogram using short-time-fourier-transform and then multiplied with the mel filterbank. The output is a two-dimensional time-frequency representation $S_{f\times t}$ input feature of $N$ is usually the spectrogram of $s$.
% The goal of the time dimension reduction system is equivalent to learning a warp matrix $W$ that converts the original spectrogram into a low-dimensional spectrogram by matrix multiplication. 
% \begin{figure}[tbp] % htbp
%   \centering
%   \includegraphics[page=3,width=0.99\linewidth]{pics/graph (4).pdf}
%   \caption{Applying DiffRes on mel-spectrogram in the audio tagging task. Here $\gamma=m/8$. DiffRes is trained end-to-end with the downstream task.}
%   \label{fig-map-improvements}
% \end{figure}

% Calculating a warp matrix $\mW$ by a monotonic weight expansion~(MWE) algorithm $w: \vs \rightarrow \mW$, where $\vs$ is the normalized version of $\vs$. 3) warping time step based on the values in $\mW$. We describe the details design of $\phi$ and $\phi$ in the next sections.

\subsection{Differentiable temporal resolution modeling}

% Figure~\ref{fig-overview} illustrates adaptive temporal resolution learning with DiffRes. We introduce frame importance estimation in Section~\ref{sec:frame-importance-estimation}, and introduce the warp matrix construction function and frame warping function in Section~\ref{sec:Monotonic-Frame-warping}. Figure~\ref{fig-example-small} shows an example of DiffRes on the mel-spectrogram.

\subsubsection{Frame importance estimation}
\label{sec:frame-importance-estimation}

% \begin{wrapfigure}{r}{3cm}
%     \small
%   \includegraphics[page=1,width=1.0\linewidth]{pics/resconv.pdf}
%     \caption{$\operatorname{ResConv1D}$}
%     \label{fig:res-conv-1d}
% \end{wrapfigure}

We design a frame importance estimation module $\mathcal{H}_{\phi}$ to decide the proportion of each frame that needs to be kept in the output, which is similar to the sample weighting operation~\citep{zhang2021learning} in previous studies.
The frame importance estimation module will output a row vector $\vs^{\prime}$ with shape $1\times T$, where the element $\vs^{\prime}_{\tau}$ is the importance score of the $\tau$-th time frame $\mX_{:,\tau}$. The frame importance estimation can be denoted as
\begin{equation}
    \label{eq-importance-estimation}
    \vs^{\prime} = \sigma(\mathcal{H}_{\phi}(\mX)),
\end{equation}
where $\vs^{\prime}$ is the row vector of importance scores, and $\sigma$ is the sigmoid function. A higher value in $\vs^{\prime}_{\tau}$ indicates the $\tau$-th frame is important for classification. We apply the sigmoid function to stabilize training by limiting the values in $\vs^{\prime}$ between zero and one. 
We implement $\mathcal{H}_{\phi}$ with a stack of one-dimensional convolutional neural networks~(CNNs)~\citep{fukushima1982neocognitron, lecun1989handwritten}. Specifically, $\mathcal{H}_{\phi}$ is a stack of five one-dimensional convolutional blocks~($\operatorname{ResConv1D}$). We design the $\operatorname{ResConv1D}$ block following other CNN based methods~\citep{shu2021joint,liu2020channel,kong2021speech}. Each $\operatorname{ResConv1D}$ has two layers of one-dimensional CNN with batch normalization~\citep{ioffe2015batch} and leaky rectified linear unit activation functions. We apply residual connection~\citep{he2016deep} for easier training of the deep architecture~\citep{zaeemzadeh2020norm}. 
% The kernel size of each CNN is five, and e
Each CNN layer is zero-padded to ensure the temporal dimension does not change~\citep{lecun2015deep}. 
We use exponentially decreasing channel numbers to reduce the computation. 
In the next frame warping step~(Section~\ref{sec:Monotonic-Frame-warping}), elements in the importance score will represent the proportion of each input frame that contributes to an output frame. Therefore, we perform rescale operation on $\vs^{\prime}$, resulting in an $\vs$ that satisfies $\vs\in[0,1]^{1\times T}$ and $\sum_{k=1}^{T}\vs_{k} \leq t$.
% $\sum^{M}_{t=1}\vs_{t} \leq \gamma$ 
% and $\text{max}(\vs)\leq 1$. 
The rescale operation can be denoted as $\check{\vs}=\frac{\vs^{\prime}}{\sum^{T}_{i=1}\vs_{i}^{\prime}}t,~~\vs=\frac{\check{\vs}}{\operatorname{max}(\check{\vs},1)}$,
% \begin{equation}
%     \label{eq-normalize}
% \end{equation}
where $\check{\vs}$ is an intermediate variable that may contain elements greater than one, $\operatorname{max}$ denotes the maximum operation. To quantify how active $\mathcal{H}_{\phi}$ is trying to distinguish between important and less important frames, we also design a measurement, activeness~$\rho$, which is calculated by the standard derivation of the non-empty frames, given by
\begin{equation}
    \label{eq-activeness}
    \rho = \frac{1}{1-\delta}\sqrt{\frac{\sum_{i\in\mathbb{S}_{\text{active}}}(\vs_{i}-\bar{\vs}_{i})^{2}}{|\mathbb{S}_{\text{active}}|}},
    % ~\delta=\frac{t-\gamma}{t},
    % ~\mu=\frac{\sum_{i\in\mathbb{S}}\vs_{i}}{|\mathbb{S}|},
    % \mu=\overline{\vs_{i}},~~
\end{equation}
\begin{equation}
    \mathbb{S}_{\text{active}}=\{i~|~\operatorname{E}(\mX_{:,i}) > \operatorname{min}({\operatorname{E}(\mX_{:,i}}))+\epsilon \},
\end{equation}
where $\mathbb{S}_{\text{active}}$ is the set of indices of non-empty frames, $\epsilon$ is a small value, $|\mathbb{S}|$ denotes the size of set $\mathbb{S}$, function $\operatorname{E}(\cdot)$ calculates the root-mean-square energy~\citep{law2015dictionary} of a frame in the spectrogram, and function $\operatorname{min}(\cdot)$ calculates the minimum value within a matrix. We use $\delta$ to unify the value of $\rho$ for easier comparison between different $\delta$ settings. The activeness $\rho$ can be used as an indicator of how DiffRes behaves during training. A higher $\rho$ indicates the model is more active at learning the frame importance. A lower $\rho$ such as zero indicates learning nothing. We will discuss the learning process of DiffRes with $\rho$ in Section~\ref{sec:final-analysis}.

\subsubsection{Temporal Frame Warping}
\label{sec:Monotonic-Frame-warping}

\begin{figure*}[t]
\centering
  \includegraphics[page=1,width=1.0\linewidth]{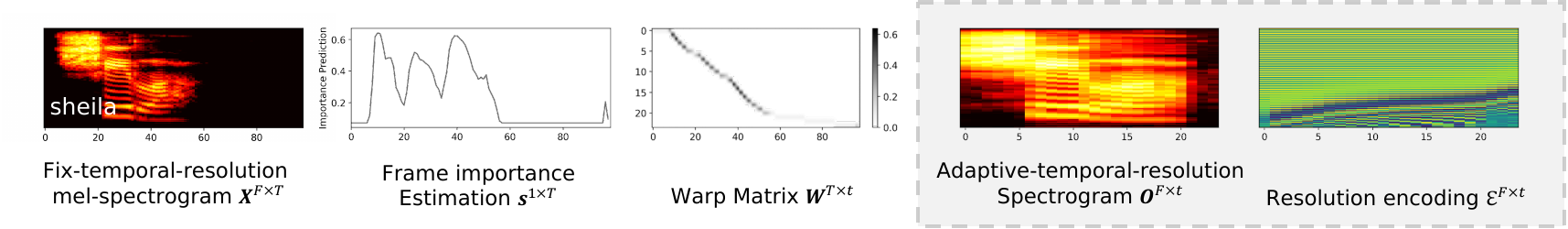}
    \caption{Visualizations of the DiffRes using the mel-spectrogram. The part with the shaded background is the input features.}
    % For more examples please refer to Figure~\ref{fig-example-sc} and~\ref{fig-example-audioset} in \hl{Appendix}~\ref{app-examples}.
    \label{fig-example-small}
\end{figure*}

We perform temporal frame warping based on $\vs$ and $\mX$ to calculate an adaptive temporal resolution representation $\mO$, which is similar to the idea of generating derived features~\citep{pentreath2015machine}. 
% Specifically, the algorithm will merge the adjacent frames based on $\vs$. 
% The word \textit{monotonic} means the output feature of the warping algorithm should preserve the original temporal order. Here monotonicity is a strong inductive bias~\citep{mitchell1980need, gordon1995evaluation} in the DiffRes algorithm considering the audio signal does not transmit in reverse time. The monotonicity makes the DiffRes feature more interpretable and performs better on the audio signal. 
Generally, the temporal frame warping algorithm can be denoted by $\mW=\alpha(\vs)$ and $~\mO=\beta(\mX, \mW)$,
% \begin{equation}
%     \label{eq-monotonic-frame-warping}
% \end{equation}
where $\alpha(\cdot)$ is a function that convert $\vs$ into a warp matrix $\mW$ with shape $t \times T$, and $\beta(\cdot)$ is a function that applies $\mW$ to $\mX$ to calculate the warpped feature $\mO$. Elements in $\mW$ such as $\mW_{i,j}$ denote the contribution of the $j\text{-th}$ input frame $\mX_{:,j}$ to the $i\text{-th}$ output frame $\mO_{:,i}$. We will introduce the realization of $\alpha(\cdot)$ and $\beta(\cdot)$ as follows. 

% Figure~\ref{fig-fast_algorithm} in \hl{Appendix}~\ref{app-sum-to-one} provides an example of the warp matrix construction function.

Function~$\alpha(\cdot)$ calculates the warp matrix $\mW$ with $\vs$ by:
\begin{equation}
    \label{eq:initial-weight}
    \mW_{i,j}=
    \begin{cases}
      \vs_{j}, & \text{if}~ i < \sum_{k=1}^{j}\vs_{k} \leq i+1 \\
      0, & \operatorname{otherwise}
    \end{cases},
  \end{equation}
where we calculate the cumulative sum of $\vs$ to decide which output frame each input frame will be warped into.
The warp matrix $\mW$ will be used for frame warping function $\beta(\cdot)$. 
Function $\beta(\cdot)$ performs frame warping based on the warp matrix $\mW$. The $i{\text{-th}}$ output frame is calculated with $\mX$ and the $i{\text{-th}}$ row of $\mW$, given by 
\begin{equation}
    \label{eq:1-frame-warping}
    \mO_{i,j} = \mathcal{A}((\mX_{j, :}) \odot (\mW_{i,:})),
\end{equation}
where $\mathcal{A}: \mathbb{R}^{1\times T} \rightarrow \mathbb{R}$ stands for the frame aggregation function such as averaging, $\mO$ is the final output feature with shape $F\times t$. 

\noindent\textbf{Resolution Encoding.}~ The final output $\mO$ does not contain the resolution information at each time step, which is crucial information for the classifier. Since the temporal resolution can be represented with $\mW$,
we construct a resolution encoding with $\mW$ in parallel with frame warping. Firstly, we construct a positional encoding matrix $\mE$ with shape $F\times T$, using the similar method described in~\citet{vaswani2017attention}.
Each column of $\mE$ represents a positional encoding of a time step. Then we calculate the resolution encoding by $\mathcal{E}=\mE\mW^{\top}$, where $\mW^{\top}$ stands for the transpose of $\mW$. The shape of the resolution encoding is $F\times t$. 
Both $\mathcal{E}$ and $\mO$ are concatenated on the channel dimension as the classifier input feature.

\subsection{Optimization}
\label{sec:end-to-end-optimization}

We propose a guide loss to provide guidance for DiffRes on learning frame importance. Since we do not know the ground truth frame importance, we cannot directly optimize $\vs$. We introduce $\mathcal{L}_{guide}$ as an inductive bias~\citep{mitchell1980need} to the system based on the assumption that an empty frame should have a low importance score. 
Specifically, we propose the guide loss by
\begin{equation}
    \label{eq:guide-loss}
    \mathcal{L}_{guide} = \frac{1}{|\mathbb{S}_{\text{empty}}|}\sum_{i \in \mathbb{S}_{\text{empty}}} (\frac{\vs_{i}}{1-\delta}-\lambda)^{+},
\end{equation}
\begin{equation}
    \mathbb{S}_{\text{empty}}=\{i~|~i \notin \mathbb{S}_{\text{active}} ~\operatorname{and}~ i\in \{1,2,...,T\} \},
\end{equation}
where $\mathbb{S}_{\text{empty}}$ is a set of time indexes that have low energy, and $\lambda$ is a constant threshold. Given that the output of DiffRes has fewer temporal dimensions than $\mX$, the DiffRes layer forms an information bottleneck~\citep{tishby2000information, shwartz2017opening} that encourages DiffRes to assign a higher score to important frames. We analyze the information bottleneck effect of DiffRes in Section~\ref{sec:final-analysis}. The parameter $\lambda$ is a threshold for the guide loss to take effect. This threshold can alleviate the modeling bias toward energy. For example, if $\lambda=0$, the importance scores of empty frames are strongly regularized, and the model will also tend to predict low importance scores for lower energy frames, which may contain useful information. $\mathcal{L}_{bce}$
is the standard binary cross entropy loss function ~\citep{shannon2001mathematical} for classification, given by Equation~\ref{eq-bce-loss}, where $\hat{\vy}$ is the label prediction and $N$ is the total number of classes.
 \begin{equation}
    \label{eq-bce-loss}
     \mathcal{L}_{bce} =\frac{1}{N}\sum_{i=1}^{N}(\vy_{i}\operatorname{log}(\hat{\vy}_{i})+(1-\vy_{i}\operatorname{log}(1-\hat{\vy}_{i}))),
 \end{equation}

The loss function of the DiffRes-based audio classification system includes our proposed guide loss $\mathcal{L}_{guide}$ and the binary cross entropy loss $\mathcal{L}_{bce}$, given by $\mathcal{L} = \mathcal{L}_{bce}+\mathcal{L}_{guide}$.

\section{Experiments}
\label{sec:experiment}
We focus on evaluating DiffRes on the mel-spectrogram, which is one of the most popular features used by state-of-the-art systems~\citep{chen2022hts, gong2022ssast, koutini2021efficient}. We evaluate DiffRes on five different tasks and datasets including audio tagging on AudioSet~\citep{gemmeke2017audio} and FSD50K~\citep{fonseca2021fsd50k}, environmental sound classification on ESC50~\citep{piczak2015esc}, limited-vocabulary speech recognition on SpeechCommands~\citep{warden2018speech}, and music instrument classification on NSynth~\citep{engel2017neural}. All the datasets are resampled at a sampling rate of $16$~kHz. Following the evaluation protocol in the previous works~\citep{zeghidour2021leaf,riad2021learning, kong2020panns,gong2021psla}, we report the mean average precision~(mAP) as the main evaluation metric on AudioSet and FSD50K, and report classification accuracy~(ACC) on other datasets. In all experiments, we use the same architecture as used by ~\citet{gong2021psla}, which is an EfficientNet-B2~\citep{tan2019efficientnet} with four attention heads~($13.6$~M parameters). We reload the ImageNet pretrained weights for EfficientNet-B2 in a similar way to~\citep{gong2021ast, gong2021psla}. For the training data, we apply random spec-augmentation~\citep{park2019specaugment} and mixup augmentation~\citep{zhang2017mixup} following~\citet{gong2021psla}.
All experiments are repeated three times with different seeds to reduce randomness. We also report the standard derivation of the repeated trials along with the averaged result. We train the DiffRes layer with $\lambda=0.5$ and $\epsilon=\num{1e-4}$. For the frame aggregation function $\mathcal{A}$~(see Equation~\ref{eq:1-frame-warping}), we use both the max and mean operations, whose outputs are concatenated with the resolution encoding $\mathcal{E}$ on the channel dimension as the input feature to the classifier. The frame importance estimation module we used in this paper is a stack of five $\operatorname{ResConv1D}$ with around $82$k  parameters. 
We calculate the mel-spectrogram with a Hanning window, $25$~ms window length, $10$~ms hop size, and  $128$~mel-filterbanks by default. 
We list the implementation details and hyperparameters setting in the supplementary material.

\subsection{Adaptively compress the temporal dimension}
Compression of mel-spectrogram temporal dimension can lead to a considerable speed up on training and inference~\citep{simpf}, which has significant promise in on-device scenarios. In this section, we evaluate the effectiveness of DiffRes in compressing temporal dimensions and classification performance.  
We compare DiffRes with three temporal dimension reduction methods: i) \textit{Change hop size}~(CHSize) reduces the temporal dimension by enlarging the hop size. The output of CHSize has a fixed resolution and may lose information between output frames.; ii) \textit{AvgPool} is a method that performs average pooling on a $100$ FPS spectrogram to reduce the temporal dimensions. 
AvgPool also has a fixed resolution, but it can aggregate information between output frames by pooling; iii) \textit{ConvAvgPool} is the setting that the $100$ FPS mel-spectrogram will be processed by a stack of $\operatorname{ResConv1D}$~(mentioned in Section~\ref{sec:frame-importance-estimation}), followed by an average pooling for dimension reduction. ConvAvgPool has around $493$k parameters. Based on a learnable network, ConvAvgPool has the potential of learning more suitable features and temporal resolution implicitly. We provide detailed implementations in the supplementary material.

\begin{table*}[t]
\centering
\footnotesize
\setlength\tabcolsep{3pt}%
\begin{tabular*}{\textwidth}{@{\extracolsep{\fill}}ccccccc@{}}
\toprule[\heavyrulewidth]
\begin{tabular}[c]{@{}c@{}}Task name \\ $100$ FPS baseline~(\%)\end{tabular} & \multicolumn{1}{l}{Metric} & FPS & \begin{tabular}[c]{@{}c@{}}Change hop size~(\%)\end{tabular}           & AvgPool~(\%)                  & \begin{tabular}[c]{@{}c@{}}ConvAvgPool~(\%)\end{tabular}                   & Proposed~(\%)                   \\
\midrule 
\multirow{3}{*}{\begin{tabular}[c]{@{}c@{}}AudioSet tagging \\
$43.7\pm0.1$ \end{tabular}} &
\multirow{3}{*}{mAP} &
  $25$ &
  $38.6\pm0.3$ &
  $39.9\pm0.2$ &
  $40.1\pm0.2$ &
  $\mathbf{41.7}\pm0.1$ \\ &
         & $50$  & $41.8\pm0.2$ & $42.4\pm0.1$ & $42.7\pm0.2$ & $\mathbf{43.6}\pm0.1$ \\ &
         & $75$  & $42.7\pm0.2$ & $43.6\pm0.0$ & $43.5\pm0.2$ & $\mathbf{44.2}\pm0.1^{\dag}$ \\
\midrule 
\multirow{3}{*}{\begin{tabular}[c]{@{}c@{}}FSD50K tagging \\ $55.6\pm0.3$\end{tabular}} &
\multirow{3}{*}{mAP} &
$25$ & $48.9\pm0.4$ & $51.4\pm0.3$ & $49.2\pm0.4$ & $\mathbf{56.9}\pm0.2^{\dag}$ \\ &
                              & $50$ & $53.3\pm0.4$ & $54.5\pm0.4$ & $52.2\pm0.8$ & $\mathbf{57.2}\pm0.2^{\dag}$ \\ &
                              & $75$ & $54.8\pm0.4$ & $55.3\pm0.3$ & $54.4\pm0.2$ & $\mathbf{57.1}\pm0.4^{\dag}$ \\
\midrule                              
\multirow{3}{*}{\begin{tabular}[c]{@{}c@{}}Environmental sound \\ $85.2\pm0.5$\end{tabular}} &
\multirow{3}{*}{ACC} &
  $25$ &
  $74.6\pm0.6$ &
  $75.6\pm0.3$ &
  $72.4\pm1.2$ &
  $\mathbf{82.9}\pm0.5$ \\ &
         & $50$  & $82.4\pm0.5$ & $83.2\pm0.3$ & $77.3\pm0.8$ & $\mathbf{85.5}\pm0.4^{\dag}$ \\ &
         & $75$  & $84.9\pm0.3$ & $85.2\pm0.4^\dag$ & $81.8\pm0.6$ &  $\mathbf{86.8}\pm0.3^{\dag}$ \\
\midrule 
\multirow{3}{*}{\begin{tabular}[c]{@{}c@{}}Speech recognition \\ $97.2\pm0.1$\end{tabular}} &
\multirow{3}{*}{ACC} &
  $25$ &
  $93.5\pm0.1$ &
  $94.9\pm0.4$ &
  $\mathbf{95.8}\pm0.3$ &
  $95.0\pm0.3$ \\ &
         & $50$  & $96.1\pm0.1$ & $96.0\pm0.2$ & $96.0\pm0.1$ & $\mathbf{96.7}\pm0.2$ \\ &
         & $75$  & $96.8\pm0.2$ & $96.9\pm0.1$ & $97.0\pm0.1$ & $\mathbf{97.2}\pm0.0^{\dag}$ \\
\midrule 
\multirow{3}{*}{\begin{tabular}[c]{@{}c@{}}Music~instrument\\ $79.9\pm0.2$\end{tabular}} &
\multirow{3}{*}{ACC} &
  $25$ &
  $79.7\pm0.2$ &
  $78.3\pm0.7$ &
  $78.0\pm0.5$ &
  $\mathbf{80.5}\pm0.2^{\dag}$ \\ &
         & $50$  & $79.9\pm0.0^{\dag}$ & $79.5\pm0.3$ & $79.4\pm0.3$ & $\mathbf{81.0}\pm0.5^{\dag}$ \\ &
         & $75$  & $79.8\pm0.2$ & $79.6\pm0.3$ & $79.7\pm0.4$ & $\mathbf{80.8}\pm0.2^{\dag}$ \\
\bottomrule
% \midrule[\heavyrulewidth]
\end{tabular*}
\caption{Comparison of different temporal dimension reduction methods. The numbers under the task name show the baseline performance. Baseline methods use fix-temporal-resolution mel-spectrogram with $10$ ms hop size. Numbers with $\dag$ mean better or comparable performance compared with the 100 FPS baseline.}
\label{tab:comparison-CHSize-avgpool-ConvAvgPool-proposed}
\end{table*}

\noindent\textbf{Baseline Comparisons. }~ Table~\ref{tab:comparison-CHSize-avgpool-ConvAvgPool-proposed} shows our experimental result. 
The baseline of this experiment is performed on mel-spectrogram without temporal compression~(i.e.,~$100$ FPS) and the baseline result is shown under each task name. When reducing $25$\% of the temporal dimension~(i.e., $75$ FPS), the proposed method can even considerably improve the baseline performance on most datasets, except on speech recognition tasks where we maintain the same performance. We assume the improvement comes from the data augmentation effect of DiffRes, which means divergent temporal compression on the same data at different training steps. With a $50$ FPS, four out of five datasets can maintain comparable performance. With only $25$ FPS, the proposed method can still improve the FSD50K tagging and music instrument classification tasks, which indicates the high temporal redundancy in these datasets. 
Our proposed method also significantly outperforms other temporal dimension reduction baselines. With fixed resolution and fewer FPS, the performance of CHSize degrades more notably. 
AvgPool can outperform CHSize by aggregating more information between output frames. 
Although ConvAvgPool has an extra learnable neural network, it does not show significant improvements compared with AvgPool. ConvAvgPool even has an inferior performance on FSD50K and environmental sound classification tasks. 
This indicates employing a simple learnable front-end for feature reduction is not always beneficial.

\noindent\textbf{On Variable-length Audio Data.}~ We observe that the proposed method improves the mAP performance by $1.3$\% with only $25$ FPS on the FSD50K dataset. We analyze it because the audio clip durations in the FSD50K have a high variance (i.e., from $0.3$ to $30$s). In previous studies~\citep{gong2021ast,gong2021psla,kong2020panns}, a common practice is padding the audio data into the same duration in batched training and inference, which introduces a considerable amount of temporal redundancy in the data with a significantly slower speed. By comparison, DiffRes can unify the audio feature shape regardless of their durations. Model optimization becomes more efficient with DiffRes. As a result, the proposed method can maintain an mAP of $55.6\pm0.2$ on the FSD50K, which is comparable to the baseline, with only $15$ FPS and $28$\% of the original training time. This result shows that DiffRes provides a new mind map for future work on classifying large-scale variable-length audio clips. 

\subsection{Learning with higher temporal resolution}
\label{sec:small_hop_size}
Previous studies have observed that a higher resolution spectrogram can improve audio classification accuracy~\citep{kong2020panns,ferraro2021low}. However, a hop size smaller than 10 ms has not been widely explored. This is partly because the computation becomes heavier for a smaller hop size. For example, with $1$~ms hop size~(i.e., $1000$ FPS), the time and space complexity for an EfficientNet classifier will be $10$ times heavier than with a common $10$~ms hop size. 
Since DiffRes can control the temporal dimension size, namely FPS, working on a small hop size spectrogram becomes computationally friendly. 
Table~\ref{tab:resolution-hop-size-test} shows model performance can be considerably improved with smaller hop sizes. AudioSet and environment sound dataset achieve the best performance on $6$~ms and $1$~ms hop size, and other tasks benefit most from $3$~ms hop sizes. In later experiments, we will use these best hop size settings on each dataset.

\begin{table}[htbp]
\centering
\footnotesize
\begin{tabular}{lcccc}
\toprule[\heavyrulewidth]
Hop size             & $10$~ms                      & $6$~ms                       & $3$~ms                       & $1$~ms                       \\
\midrule
AudioSet & $43.7\pm0.1$ & $\mathbf{44.1}\pm0.1$ & $43.8\pm0.0$ & $43.7\pm0.1$ \\
ESC-50 & $85.2\pm0.4$ & $87.2\pm0.3$ & $88.0\pm0.6$ & $\mathbf{88.4}\pm0.5$ \\
SC      & $97.2\pm0.1$ & $97.6\pm0.0$ & $\mathbf{97.9}\pm0.1$ & $97.8\pm0.1$ \\
NSynth   & $79.9\pm0.2$ & $81.3\pm0.3$ & $\mathbf{81.8}\pm0.2$ & $80.6\pm0.4$ \\
\midrule
Avg.   & $76.5\pm0.2$ & $77.6\pm0.2$ & $\mathbf{77.9}\pm0.1$ & $77.5\pm0.2$ \\
\midrule
\end{tabular}
\caption{Learning with high temporal resolution spectrograms. FPS is controlled at 100, so the computational complexity of the classifier is the same in all hop-size settings. Results are reported in the percentage format.}
\label{tab:resolution-hop-size-test}
\end{table}

\noindent\textbf{Comparing with Other Learnable Front-ends.}~ The DiffRes is differentiable, so the Mel+DiffRes setting as a whole can be viewed as a learnable front-end. Table~\ref{tab:learnable-front-end-compare} compares our proposed method with SOTA learnable front-ends, our best setting is denoted as Mel+DiffRes~(Best), which achieves the best result on all datasets. For a fair comparison, we control the experiment setup to be consistent with ~\citet{zeghidour2021leaf} in Mel+DiffRes. Specifically, we change the backbone to EfficientNet-B0~($5.3$~M parameters) without ImageNet pretraining. We also remove spec-augment and mixup, except in AudioSet, and change our Mel bins from $128$ to $40$, except in the AudioSet experiment where we change to 64. The result shows Mel+DiffRes can outperform SOTA learnable front-end~\citep{zeghidour2021leaf, ravanelli2018speaker, zeghidour2017learning} by a large margin, demonstrating the effectiveness of DiffRes.

\begin{table*}[htbp]
\centering
\small
\setlength\tabcolsep{3pt}%
\begin{tabular}{lcccccc}
\toprule[\heavyrulewidth]
   Front-end        & Mel & TD-fbank & SincNet & LEAF & Mel+DiffRes & Mel+DiffRes (Best) \\
\midrule
Parameters & $0$   & $51$~k     & $256$     & $448$  & $82$~k          & $82$~k          \\
% \midrule
% Frames per second~(FPS)        & \multicolumn{6}{c}{100}   
% \\
\midrule
AudioSet tagging &
  $96.8\pm0.1$ &
  $96.5\pm0.1$ &
  $96.1\pm0.0$ &
  $96.8\pm0.1$ &
  $97.0\pm0.0$ &
  $\mathbf{97.5}\pm0.0$ \\
Speech recognition &
  $93.6\pm0.3$ &
  $89.5\pm0.4$ &
  $91.4\pm0.4$ &
  $93.6\pm0.3$ &
  $95.4\pm0.2$ &
  $\mathbf{97.9}\pm0.1$ \\
% Music (Pitch) &
%   88.5\pm0.4 &
%   86.4\pm0.4 &
%   81.2\pm0.5 &
%   88.6\pm0.4 &
%   91.8\pm0.6 &
%   \mathbf{93.0}\pm0.4 \\
Music instrument &
  $70.7\pm0.6$ &
  $66.3\pm0.6$ &
  $67.4\pm0.6$ &
  $70.2\pm0.6$ &
  $78.5\pm0.7$ &
  $\mathbf{81.8}\pm0.2$ \\
\midrule
Average &
  $87.0\pm0.3$ &
  $84.1\pm0.4$ &
  $85.0\pm0.3$ &
  $86.9\pm0.3$ &
  $90.3\pm0.3$ &
  $\mathbf{92.4}\pm0.1$ \\
\midrule[\heavyrulewidth]
\end{tabular}
\caption{Comparison with SOTA learnable front-ends. All the methods use $100$ FPS. Results are reported in the percentage format. Mel+DiffRes controls the experimental settings mentioned in Section~\ref{sec:small_hop_size} to be consistent with Mel, TD-fbank, SincNet, and LEAF. Mel+DiffRes~(Best) use the best possible settings.}
\label{tab:learnable-front-end-compare}
\end{table*}

\begin{figure}[t]
    \centering
  \includegraphics[width=0.6\linewidth]{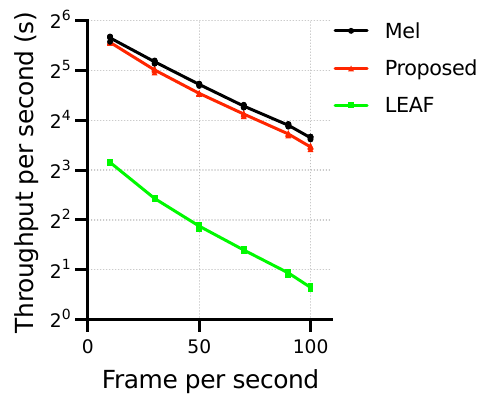}
  \captionof{figure}{Audio throughput in one second. Evaluated on a 2.6 GHz Intel Core i7 CPU. }
  \label{fig:efficiency_compare}
  \vspace{-1em}
\end{figure}

\noindent\textbf{Computational Cost.}~ We assess the one-second throughput of different front-ends on various FPS settings to compare their computational efficiency. We control the FPS of Mel and LEAF by average pooling. The computation time is measured between inputting waveform and outputting label prediction~(with EfficientNet-B2). We use $128$ filters in LEAF~\citep{zeghidour2021leaf} for a fair comparison with $128$ mel-filterbanks in Mel and DiffRes. As shown in Figure~\ref{fig:efficiency_compare}, our proposed DiffRes only introduces marginal computational cost compared with Mel. The state-of-the-art learnable front-end, LEAF, is about four times slower than our proposed method. The majority of the cost in computation in LEAF comes from multiple complex-valued convolutions, which are computed in the time-domain with large kernels~(e.g., $400$) and a stride of one.

\subsection{Analysis for the learning of DiffRes}
\label{sec:final-analysis}
\begin{figure}[tbp]
\begin{minipage}[t]{.23\textwidth}
  \centering
  \includegraphics[width=1.0\linewidth]{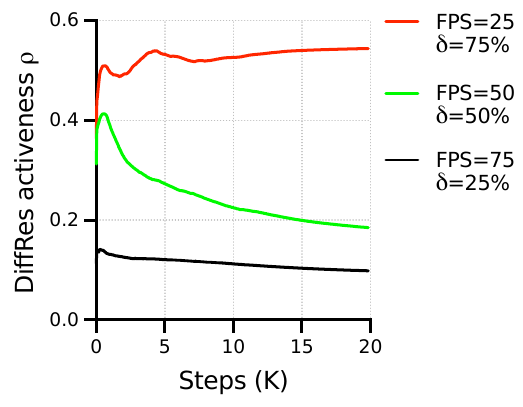}
  \captionof{figure}{Trajectories of DiffRes learning activeness~($\rho$) on different training steps and FPS settings.}
  \vspace{-1em}
  \label{fig:diffres-activeness}
\end{minipage}
\hfill
\begin{minipage}[t]{.23\textwidth}
  \centering
  \includegraphics[width=1.0\linewidth]{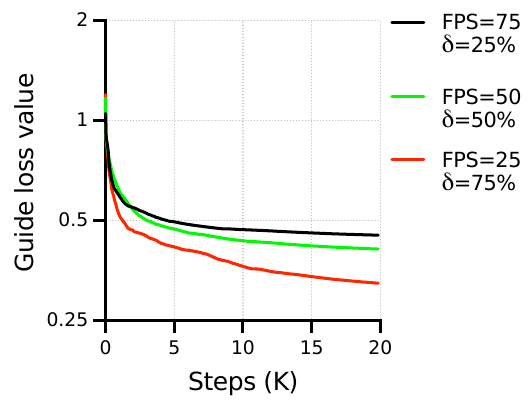}
  \captionof{figure}{The training curve of guide loss~($\mathcal{L}_{guide}$) with different FPS settings.}
  \label{fig:guide-loss-curve}
\end{minipage}
\end{figure}
\begin{figure}[htbp]
\vspace{1.5em}
  \includegraphics[page=1,width=1.0\linewidth]{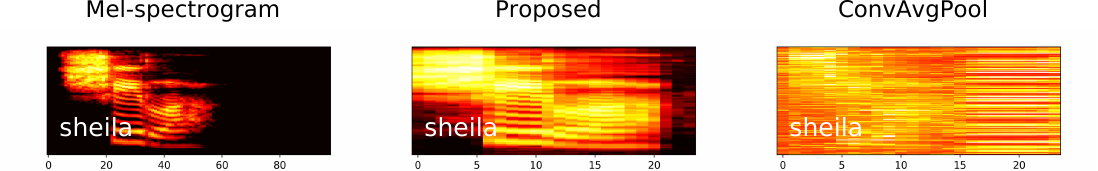}
    \caption{Comparison of mel-spectrogram, DiffRes feature, and ConvAvgPool learned feature. The DiffRes feature preserves more details in the original mel-spectrogram and is more interpretable than the ConvAvgPool feature.}
    \label{fig-interpreterbility}
    \label{app-figures}
    \vspace{-1em}
\end{figure}

% We analyze the behavior of DiffRes during optimization in this section. 
\textbf{Learning Activeness.}~ DiffRes does not explicitly learn the optimal frame importance score because the ground truth frame importance is not available. Instead, DiffRes is optimized with the guidance of guide loss $\mathcal{L}_{guide}$~(Equation~\ref{eq:guide-loss}), which is a strong assumption we introduced to the model. Figure~\ref{fig:diffres-activeness} shows the trajectories of the DiffRes learning activeness~(defined in Section~\ref{sec:frame-importance-estimation}) during the optimization with different FPS settings on the speech recognition task in Table~\ref{tab:comparison-CHSize-avgpool-ConvAvgPool-proposed}. According to the final converged value, DiffRes with a smaller FPS tends to be more active at learning frame importance. This is intuitive since smaller FPS leads to more information bottleneck effects~\citep{saxe2019information} in DiffRes. With a $25$ FPS, the activeness even keeps increasing with more training steps, indicating the active learning of DiffRes. Figure~\ref{fig:guide-loss-curve} shows the guide loss curve during training with different FPS settings. Intuitively, when the FPS is small, a model needs to preserve more non-empty frames and fewer empty frames for better accuracy. This assumption is aligned with our experiment result, which shows the model tends to have a lower guide loss with a smaller FPS. 

\noindent\textbf{Data Augmentation and Regularization Effect.}~ As reflected in the curve of $\rho$ and $\mathcal{L}_{guide}$ in Figure~\ref{fig:diffres-activeness} and~\ref{fig:guide-loss-curve}, DiffRes is optimized along with the classifier during training. Hence DiffRes produces different outputs for the same training data at different epochs. This is equivalent to performing data augmentation on the audio data. We suppose this is the main reason for the improved performance shown in Table~\ref{tab:comparison-CHSize-avgpool-ConvAvgPool-proposed}. Also, DiffRes reduces the sparsity of the audio feature by adaptive temporal compression. This is equivalent to performing an implicit regularization~\citep{neyshabur2017implicit,arora2019implicit} on the feature level, which is beneficial for the system efficiency.

\noindent\textbf{Ablation Studies.}~ We further study the effect of hyper-parameters used in the DiffRes method, including guide loss (e.g., activeness $\rho$, threshold $\lambda$), dimension reduction ratio $\delta$, small value $\epsilon$. We also experiment with different model architectures~\citep{kong2020panns}). We perform the ablation studies on the SpeechCommands dataset since it has a reasonable amount of data and is computationally friendly on model training. We also study the potential of DiffRes in multimodal tasks such as audio captioning~\citep{kim2019audiocaps}. Results and analysis of the ablation studies are shown in the supplementary materials.

\noindent\textbf{Visualization.}~ We visualize the compression results of DiffRes, as compared with the \textit{ConvAvgPool} method. The visualization results are as shown in Figure~\ref{fig-interpreterbility}. We observe that DiffRes learns to remove silent frames and compress the pattern in the mel-spectrogram. This observation shows the both effectiveness and interpretability of DiffRes.

\section{Related Work}
Neural-network based methods have been successfully applied to audio classification and achieved state-of-the-art performance, such as the pre-trained audio neural networks~(PANNs)~\citep{kong2020panns}, pretraining, sampling, labeling, and aggregation-based audio tagging~(PSLA)~\citep{gong2021psla}, and audio spectrogram transformer~(AST)~\citep{gong2021ast}. We will cover two related topics on audio classification in the following sections. 

\textbf{Learnable Audio Front-ends.}~ In recent years, learning acoustic features from waveform using trainable audio front-ends has attracted increasing interest from researchers. \cite{sainath2013learning} introduced one of the earliest works that propose to jointly learn the parameter of filter banks with a speech recognition model. 
Later, SincNet~\citep{ravanelli2018speaker} proposes to learn a set of bandpass filters on the waveform and has shown success on speaker recognition~\citep{ravanelli2018speaker,ravanelli2018interpretable}. Most recently, ~\citep{zeghidour2021leaf} proposes to learn bandpass, and lowpass filtering as well as per-channel compression~\citep{wang2017trainable} simultaneously in the audio front-end and shows consistent improvement in audio classification. Different from existing work on learnable audio front-ends, which mostly focus on the frequency dimension, our objective is learning the optimal temporal resolution. We show that our method can outperform existing audio front-ends for audio classification on both accuracy and computation efficiency~(see Table~\ref{tab:learnable-front-end-compare} and Figure~\ref{fig:efficiency_compare}). Note that our proposed method can also be applied after most learnable front-ends~\citep{zeghidour2021leaf}, which will be our future direction.

\textbf{Learning Feature Resolution.}~ 
One recent work on learning feature resolution for audio classification is DiffStride~\citep{riad2021learning}, which learns stride in convolutional neural network~(CNN) in a differentiable way and outperforms previous methods using fixed stride settings. By comparison, DiffStride needs to be applied in each CNN layer and can only learn a single fixed stride setting, while DiffRes is a one-layer lightweight algorithm and can personalize the best temporal resolution for each audio during inference. Recently, \cite{gazneli2022end} proposed to use a stack of one-dimension-CNN blocks to downsample the audio waveform before the audio classification backbone network, e.g., Transformer, which can learn temporal resolution implicitly for audio classification. 
In contrast, DiffRes can explicitly learn temporal resolution on the feature level with similar interpretability as the mel-spectrogram. 

\section{Conclusions}
% The proposed algorithm can integrate with any time-frequency representation at a negligible computational cost. 

In this paper, we introduce DiffRes, a ``drop-in'' differentiable temporal resolution learning module that can be applied between audio spectrogram and downstream tasks. For the training of DiffRes, our proposed guide loss is shown to be beneficial. We demonstrate over a large range of tasks that DiffRes can improve or maintain similar performance with $25$\% to $75$\% reduction on temporal dimensions, and can also efficiently utilize the information in high-resolution spectrograms to improve accuracy. In future work, we will move forward to evaluate DiffRes on different kinds of time-frequency representations with more sophisticated frame importance prediction models. Also, we will explore the potential of DiffRes in other time series data as well for learning optimal temporal resolutions.

\section{Acknowledgement}
This research was partly supported by the British Broadcasting Corporation Research and Development~(BBC R\&D), Engineering and Physical Sciences Research Council (EPSRC) Grant EP/T019751/1 ``AI for Sound'', and a PhD scholarship from the Centre for Vision, Speech and Signal Processing (CVSSP), Faculty of Engineering and Physical Science (FEPS), University of Surrey. For the purpose of open access, the authors have applied a Creative Commons Attribution (CC BY) license to any Author Accepted Manuscript version arising. 

\bibliography{aaai24}

\end{document}

% \title{Technical Appendix}

\begin{table*}[t]
\small
\centering
\begin{tabular}{cccccc}
\toprule[\heavyrulewidth]
Task &
  \begin{tabular}[c]{@{}c@{}}Audio \\ tagging\end{tabular} &
  \begin{tabular}[c]{@{}c@{}}Audio \\ tagging\end{tabular} &
  \begin{tabular}[c]{@{}c@{}}Environmental \\ sound\end{tabular} &
  \begin{tabular}[c]{@{}c@{}}Speech \\ recognition\end{tabular} &
  \begin{tabular}[c]{@{}c@{}}Music \\ instrument \end{tabular} \\
\midrule
Dataset        & AudioSet & FSD50K & ESC50 & SpeechCommands & NSynth  \\
\midrule
Classes        & $527$      & $200  $  & $50 $   & $35  $           & $11$    \\
Train examples & $1912134$  & $36799$  & $2000$  & $84771$        & $289205$  \\
Test examples  & $18887$    & $10231$  & -     & $10700$          & $12678$  \\
Duration~(mean,std)  & $9.91, 0.50$     & $7.63, 7.82 $  & $5.00, 0.00$  & $0.98,0.07  $         & $4.00,0.00$    \\
Pad to length   & $1000$     & $3000 $  & $500$  & $98$          & $400$  \\
Evaluation metric   & mAP     & mAP  & Accuracy  & Accuracy          & Accuracy  \\
5-fold cross-validation   & -     & -  & $\checkmark$  & -          & -  \\
Class re-balancing   & $\checkmark$     & $\checkmark$  & -  & -          & -  \\
SpecAug   & $\checkmark$     & $\checkmark$  & $\checkmark$  & $\checkmark$          & $\checkmark$  \\
\midrule[\heavyrulewidth]
\end{tabular}
\caption{Detailed information of the datasets we used in this paper. We perform padding to unify the data length. The last row shows the mel-spectrogram temporal dimension we used for batched training.}
\label{tab:dataset-info}
\end{table*}

\section{Ablation Studies}

We report ablation studies on hyper-parameters and different model architectures, respectively, as mentioned in Section 3.3 of the main paper. We also discuss whether DiffRes only learns to remove silent frames.

\subsubsection{Hyper-parameters}
\label{appendix-hyper-parameters}
In this section, we provide ablation studies and discussions on the hyper-parameters in DiffRes, including the threshold $\lambda$ mentioned in Equation 8, dimension reduction rate $\delta$, and the $\epsilon$ used in Equation 4 and Equation 5. We choose to conduct the experiment on the SpeechCommands dataset since it has a reasonable amount of data and is computationally friendly on model training compared with large datasets such as AudioSet~\citep{gemmeke2017audio}. 
% Although we only experiment on one dataset, we believe the conclusions are generalizable across other datasets. 
The ablation study results on hyper-parameter is presented in Table~\ref{tab-ablation-study-hyper-param} in the supplementary material.

% Please add the following required packages to your document preamble:
% \usepackage{booktabs}
% \usepackage{multirow}
\newcommand{\cmark}{\ding{51}}%
\newcommand{\xmark}{\ding{55}}%

\begin{table*}[htbp]
\small
\centering
\begin{tabular}{@{}ccc|ccccc|c@{}}
\toprule
\multicolumn{1}{c}{Metric} & Hop size~(ms) & $\delta$                   & $\lambda\vequalsignnospace$0.0  & $\lambda\vequalsignnospace0.3$  & $\lambda\vequalsignnospace0.5$  & $\lambda\vequalsignnospace0.8$  & Average  & /        \\ \midrule
\multicolumn{1}{c}{\multirow{3}{*}{Accuracy (\%)}}                 & $3$  & 70\%  & $98.0$ & $97.9$ & $98.0$ & $98.0$ & $98.0 \pm 0.0$ & $97.8$  \\ 
\multicolumn{1}{c}{}       & $1$      & 90\%                        & $98.0$ & $98.0$ & $98.0$ & $98.0$ & $98.0$ $\pm$ $0.0$ & $97.8$ \\ 
\multicolumn{1}{c}{}       & $0.5$      & 95\%                     & $97.9$ & $97.9$ & $97.9$ & $98.0$ & $97.9$ $\pm$ $0.0$ & $97.7$ \\
\midrule
\multicolumn{1}{c}{\multirow{3}{*}{Activeness $\rho$ (\%)}}               & $3$ & 70\%  & $32.4$ & $29.4$ & $28.1$ & $30.6$ & $30.1$ $\pm$ $1.6$ & $20.6$ \\ 
\multicolumn{1}{c}{}       & 1     & 90\%                         & $42.3$ & $42.8$ & $43.9$ & $42.4$ & $42.9$ $\pm$ $0.6$ & $17.9$ \\ 
\multicolumn{1}{c}{}       & $0.5$     & 95\%                     & $43.1$ & $41.6$ & $44.7$ & $40.5$ & $42.5$ $\pm$ $1.6$ & $8.4$   \\ \midrule
\multicolumn{1}{c}{\multirow{3}{*}{\begin{tabular}[c]{@{}c@{}} Average \\ importance scores  \\ on empty frames (\%)\end{tabular}}} & $3$ & 70\%  & $0.2$  & $13.2$ & $27.2$ & $41.0$ & $20.4$ $\pm$ $17.6$ & $81.1$ \\
\multicolumn{1}{c}{}       & $1$     & 90\%                      & $0.2$  & $4.6$  & $11.8$ & $30.5$ & $11.8$ $\pm$ $13.4$ & $91.6$    \\
\multicolumn{1}{c}{}       & $0.5$   & 95\%                     & $0.2$  & $3.0$  & $11.0$ & $17.2$ & $7.8$ $\pm$ $7.7$  & $102.3$   \\ 
\midrule
\multicolumn{3}{c|}{Guide loss applied}          & \cmark & \cmark& \cmark& \cmark & \cmark  & \xmark \\ \bottomrule
\end{tabular}
\caption{Ablation study on the SpeechCommands dataset. All the experiments use $100$ FPS. The baseline performance is $97.2\pm0.1$ with $10$ ms hop size and $\delta=0\%$. We report the accuracy, activeness, and average importance score on empty frames on different hop size, dimension reduction rate $\delta$, guide loss, and threshold $\lambda$ settings. The column ``Average'' denotes the average result on each metric with four different $\lambda$ values. }
\label{tab-ablation-study-hyper-param}
\end{table*}

\noindent\textbf{The Effect of Guide Loss.~} Table~\ref{tab-ablation-study-hyper-param} in supplementary material shows that even without guide loss, the model can still improve over the baseline performance~($97.2\pm0.1$) using DiffRes. At the same time, applying guide loss can further improve the activeness $\rho$~(see Equation 4 in the main paper) of DiffRes and classification performance. 
For example, without the guide loss, the $\rho$ with $3$~ms, $1$~ms, and $0.5$~ms hop size are $20.6$, $17.9$, and $8.4$, respectively, while after applying guide loss, the average $\rho$ become $32.6$, $45.4$, and $45.0$, respectively.
The improvement on $\rho$ indicates guide loss can encourage the model to better discriminate between the importance of frames. The classification accuracy can improve by about $0.2\%$ 
after applying guide loss, which is significant enough for the SpeechCommands dataset. Moreover, without guide loss, the model tends to predict high-importance scores on empty frames, which is also counterintuitive.
% In theory, if the bottleneck effect of DiffRes is strong enough, the model will automatically learn the best important score by itself. In order to study when do we need guided loss, we conduct the following experiment.

% 1. The activeness without guided loss with different hop size settings.
% 2. The activeness with guided loss with different lambda settings.
% 3. The activeness with guided loss and fix lambda setting with different eps settings.

\noindent\textbf{The Effect of Dimension Reduction Rate $\delta$.~} With the same hop size, a smaller $\delta$ will lead to a larger temporal dimension in the DiffRes output feature, which also leads to heavier computational cost~(see Figure~4 in the main paper). Even though a smaller hop size and smaller $\delta$ tend to achieve better performance because finer temporal details can be preserved, in practice, the exact value of $\delta$ still should be determined by the computation limit.

\noindent\textbf{The Effect of Guide Loss Threshold $\lambda$.~} As shown in Table~\ref{tab-ablation-study-hyper-param} in supplementary material, we tried different $\lambda$ on different hop sizes. The experimental result shows model accuracy is not sensitive to $\lambda$ thus the value of $\lambda$ usually does not need careful finetuning.

% . As the same time, a smaller $\lambda$ will lead to smaller importance score on empty frames, which is aligned with the intuition behind the guide loss. In all, our result indicates $\lambda$ only have a minor effect on model performance

\noindent\textbf{The Effect of the Small Value $\epsilon$.~} We use $\epsilon$ in Equation 5 in the main paper to control the threshold of deciding whether each frame is active or empty. In practice, we will apply SpecAug~\citep{park2019specaugment} on the spectrogram, in which the empty frames $\mathbb{S}_{\textrm{empty}}$ in Equation 8 in the main paper will be the masked time steps. To verify $\epsilon$ is not essential for model training, We try to construct $\mathbb{S}_{\textrm{empty}}$ on training data with five different $\epsilon$ values between \num{1e-4} and \num{1e-8}. Our result shows more than 98\% training data have the same $\mathbb{S}_{\textrm{empty}}$ with different $\epsilon$ values, which indicates $\epsilon$ is not an essential hyper-parameter for model training.

\subsubsection{Model architecture}

To verify the generality of the proposed approach, we also conduct experiments on two more state-of-the-art architectures, CNN6 and CNN14~\citep{kong2020panns}. Experiments are conducted on the SpeechCommands dataset with the same setting as Table 1 in the main paper.

\label{appendix-model-architecture}
\begin{table*}[t]
\small
\centering
\begin{tabular}{ccc|ccc}
\toprule
Hop Size (ms) & $\delta$ & Frames per second & EfficientNet-b2       & CNN6                  & CNN14                \\
\midrule
$10$              & $0\%$                      & $100$ & $97.2$$\pm$$0.1$            & $96.4$ $\pm$ $0.1$          & $97.9$ $\pm$ $0.1$          \\
\midrule
$10$              & $25\%$                     & $75$  & $97.2$ $\pm$ $0.0$          & $96.4$ $\pm$ $0.0$          & $98.0$ $\pm$ $0.0$          \\
$10$              & $50\%$                     & $50$  & $96.7$ $\pm$ $0.2$          & $96.1$ $\pm$ $0.1$          & $97.7$ $\pm$ $0.0$          \\
$10$              & $75\%$                   & $25$  & $95.0$ $\pm$ $0.3$          & $95.7$ $\pm$ $0.1$          & $97.1$ $\pm$ $0.1$          \\
\midrule
$6$               & $40\%$                    & $100$ & $97.6$ $\pm$ $0.0$          & $96.8$ $\pm$ $0.0$          & \textbf{$98.1$ $\pm$ $0.0$} \\
$3$               & $70\%$                    & $100$ & \textbf{$97.9$ $\pm$ $0.1$} & \textbf{$97.2$ $\pm$ $0.1$} & \textbf{$98.1$ $\pm$ $0.0$} \\
$1$               & $90\% $                   & $100$ & $97.8$ $\pm$ $0.1$          & \textbf{$97.2$ $\pm$ $0.0$} & \textbf{$98.1$ $\pm$ $0.2$} \\
\bottomrule
\end{tabular}
\caption{Ablation study on the model architectures. We use $\delta$ to denote the dimension reduction rate. Large $\delta$ indicates less computational cost.}
\label{tab-ablation-arch}
\end{table*}

Table~\ref{tab-ablation-arch} in the supplementary material shows our ablation study results on different architectures. Three results exhibit a similar trend as Table 1 and Table 2 in the main paper. All three models can maintain a similar or better performance after reducing $25\%$ of the temporal dimensions. With the same number of frames per second, namely the same computational cost, all the models show clear improvement with a smaller hop size. This improvement indicates DiffRes is effective in selecting informative frames across different architectures. We do not experiment with other non-neural architecture because the optimization of DiffRes requires gradient back-propagations~\citep{lecun2015deep}.

\subsubsection{Remove empty frame or select important frame?}
\label{appendix-remove-empty-frames}
% The importance score can tell the truth.
% \textbf{Does DiffRes only learn to remove empty frames?}
To study whether DiffRes learns to remove only silent frames, or if it would be also effective when the signal has consistent energy, we design a pitch classification experiment on the NSynth dataset following~\citep{zeghidour2021leaf}. We will refer to this task as NSynth-Pitch. We design the pitch classification task for the following two reasons: (i) The data in NSynth is mostly instrumental sounds, which have stable spectral patterns and are highly redundant for the pitch classification task. Thus NSynth-Pitch is an ideal use case of DiffRes. (ii) Our statistic shows about $19.7\%$ frames in this dataset are silent frames, thus any dimension reduction rate $\delta$ larger than $19.7\%$ means DiffRes need to remove some non-empty frames to benefit classification accuracy.
% To address this concern we design the following experiment. 1. There are about $19.7\%$ silent frames in the nsynth-pitch dataset, and 2. the audio pattern is relatively stable. So in this dataset the redundency is big but the silent frames is not prominent. In this case we can find out if our model learn to select useful frame or simply remove the silent frames.

\begin{table*}[t]
\centering
\small
\begin{tabular}{cccc}
\toprule
Frames per second~/~Dimension reduction rate                          & $25/\delta=75\%$ & $50/\delta=50\%$                       & $75/\delta=25\%$                       \\
\midrule
AvgPool                      & $90.5 \pm 0.3$                     & $91.3 \pm 0.2$                     & $92.6 \pm 0.2$                     \\
\midrule
\multicolumn{1}{c}{Proposed} & \multicolumn{1}{l}{$92.1 \pm 0.1$} & \multicolumn{1}{l}{$92.4 \pm 0.2$} & \multicolumn{1}{l}{$92.6 \pm 0.1$} \\
\midrule
\end{tabular}
\caption{Experiment result on the pitch classification task on the NSynth dataset. All the experiments use a $10$ ms hop size. The baseline performance is $92.5\pm0.2$, with $10$ ms hop size and $100$ FPS.}
\label{tab-nsynth-pitch-ablation}
\end{table*}

Table~\ref{tab-nsynth-pitch-ablation} in the supplementary material shows our result on the NSynth-Pitch task. All the settings use a dimension reduction rate $\delta > 19.7\%$, which means DiffRes have to remove part of the non-empty frames. If we reduce the temporal dimension with AvgPool, the performance will degrade significantly, while our proposed method can remain similar performance even after reducing $75\%$ temporal dimensions. This result suggests DiffRes not only remove the silent frames but also preserves important frames for classification. The high activeness $\rho$~(see Equation 4 in the main paper) in the non-empty frames shown in Figure 5 in the main paper and Table~\ref{tab-ablation-study-hyper-param} in the supplementary material also indicate the model has learned to distinguish the importance of the non-empty frames.

\section{Automatic Audio Captioning with DiffRes}

To further verify the generality of our proposed method, we conduct an extra set of experiments on the automatic audio captioning~(AAC) task~\citep{mei2022automated}, which can automatically generate natural language descriptions for audio clips. We use the architecture proposed by~\citet{mei2021encoder} and the same DiffRes setting for the experiments. Our experiments are on the AudioCaps dataset. We will try to reduce the input feature size of the AAC task and observe the change in model performance.

We conduct experiments on AudioCaps \citep{kim2019audiocaps}, which is the largest public audio captioning dataset with around $50000$ $10$-second audio clips, and is divided into three splits: training, validation and testing sets. The audio clips are annotated by humans through the Amazon Mechanical Turk (AMT) crowd-sourced platform. Each audio clip in the training sets has a human-annotated caption, while each clip in the validation and test set has five ground-truth captions.

% Because some video clips are no longer available on YouTube, our downloaded version contains $47745$ audio clips in the training set, $480$ clips in the validation set and $928$ clips in the test set.

For model evaluation, we use the metrics calculated based on $n$-gram matching ($n$-gram refers to $n$ consecutive words) following previous works~\citep{liu2022leveraging, liu2022visually}. BLEU$_n$ measures the precision of $n$-gram matching and a sentence-brevity penalty is introduced to penalize short sentences. ROUGE$_l$ calculates an F-measure by considering the longest common subsequence between the candidate and ground truths. METEOR calculates uni-gram precision and recall, taking into account the surface forms, stemmed forms, and meanings of words. CIDEr computes the cosine similarity of weighted $n$-grams between candidates and references. SPICE parses each caption into scene graphs and an F-measure is calculated based on the matching of the graphs. SPIDEr is the average of SPICE and CIDEr and is used as the official ranking metric in DCASE challenge~\citep{mei2022automateddcase}.

\begin{table*}[htbp]
\centering
\small
\begin{tabular}{cccccccc}
\toprule
\begin{tabular}[c]{@{}c@{}} $\delta $~($\approx$FLOPs reduction)\end{tabular} &
  BLEU-1 &
  BLEU-4 &
  METEOR &
  ROUGE$_l$ &
  CIDEr &
  SPICE &
  SPIDEr \\
\midrule
$0\%$ & $0.658$ & $0.235$ & $\mathbf{0.232}$ & $0.473$ & $0.643$ & $0.168$ & $0.406$ \\
$25\%$ & $0.665$ & $0.247$ & $0.228$ & $0.471$ & $\mathbf{0.657}$ & $\mathbf{0.171}$ & $\mathbf{0.414}$ \\
$50\%$ & $\mathbf{0.674}$ & $\mathbf{0.266}$ & $0.230$ & $0.475$ & $0.646$ & $0.167$ & $0.407$ \\
$75\%$ & $0.659$ & $0.252$ & $0.225$ & $\mathbf{0.478}$ & $0.649$ & $0.164$ & $0.407$ \\
\midrule
\end{tabular}
\caption{Applying DiffRes on the automatic audio captioning task, which exhibits a similar trend with audio classification tasks shown in Table~1 and Table~2 in the main paper. By removing $25$\% of the input dimensions, the performance on some metrics even got improved. After removing $75$\% of the input temporal dimensions with DiffRes, the model can still retain a comparable result.}
\label{tab:result_captioning}
\end{table*}

The result in Table~\ref{tab:result_captioning} shows that applying DiffRes on the AAC task can significantly reduce the computational cost while preserving similar performance on most of the metrics.
We perform experiments with four different temporal dimension reduction rate settings, including $0\%$, $25$\%, $50$\%, and $75$\% reductions. The reduction on the temporal dimension also significantly benefits model throughput at the same time~(see Figure~4 in the main paper). After removing $25\%$ temporal dimensions, the performance of AAC even shows an improvement, which might be due to the data augmentation effect mentioned in main paper Section~3.3. After removing $75\%$ of the input temporal dimensions, the model can still achieve on-par results compared with the baseline $0\%$ reduction. The $75\%$ reduction setting can even improve five metrics out of the total seven metrics. The result of the AAC task further indicates our proposed method is generalizable to other similar audio tasks.

\subsection{Figures}

This section list figures that assist the comprehension of the main paper. Figure 1 in this section visually compares the feature learned by our proposed method and our baseline ConvAvgPool. Figure 2 and Figure 3 illustrate the result of our pioneering studies.

\begin{figure}[htbp]
\centering
  \includegraphics[page=1,width=0.9\linewidth]{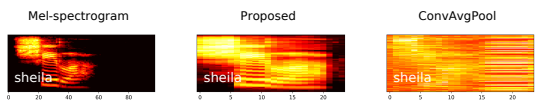}
    \caption{Comparison of mel-spectrogram, DiffRes feature, and ConvAvgPool learned feature. The DiffRes feature preserves more details in the original mel-spectrogram and is more interpretable than the ConvAvgPool feature.}
    \label{fig-interpreterbility}
    \label{app-figures}
\end{figure}

\begin{figure}[htbp]
\centering
  \includegraphics[page=1,width=0.9\linewidth]{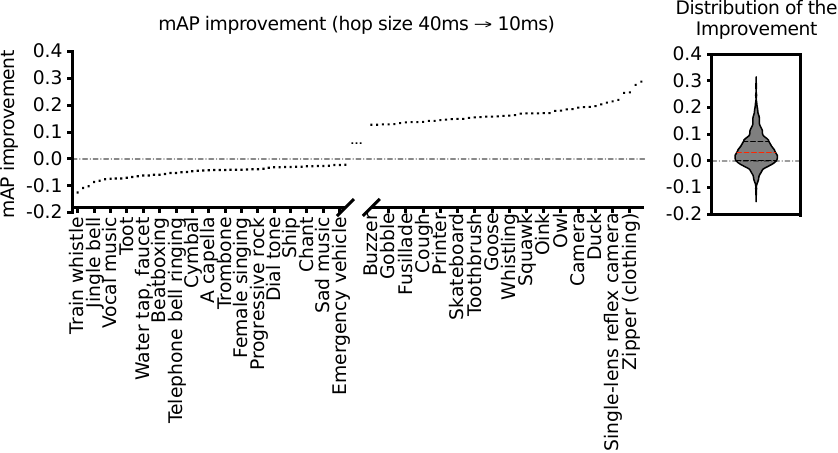}

\caption{Class-wise improvement after changing hop size from $40$~ms to $10$~ms. The mAP improvement for each class in the AudioSet after decreasing the hop size from $40$~ms to $10$~ms. The violin plot on the right side shows the improvement distribution, where the red dashed line is the median value. The inconsistency of improvement in different sound classes indicates they need different temporal resolutions to achieve optimal classification performance.}
\label{fig-map-improvements}
\end{figure}

\begin{figure}[htbp]
\centering
  \includegraphics[page=1,width=0.9\linewidth]{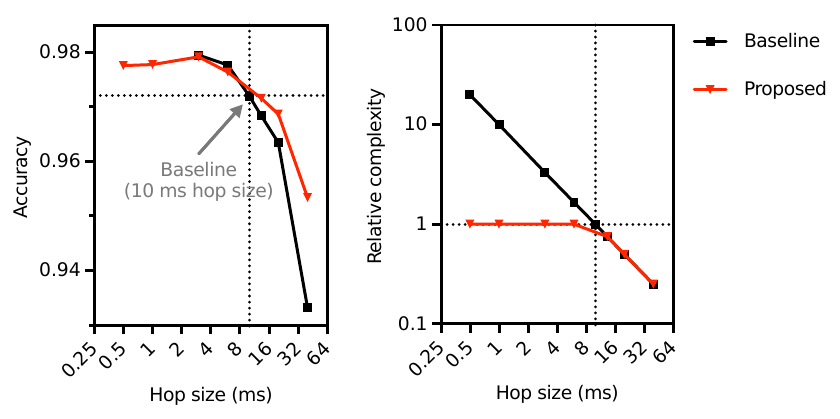}
\caption{Accuracy and the classifier computational complexity with different hop size settings on the speech recognition task. The black dotted lines show the accuracy, and complexity with a $10$~ms hop size. The accuracy can be improved with a smaller hop size at the cost of computation. DiffRes can achieve similar improvements without increasing computational complexity.} 
\label{fig-sc}
\end{figure}

\newpage

\subsection{Weakness analysis}
\label{app-weakness}
% We perform ablation studies on audio tagging tasks with AudioSet~\citep{gemmeke2017audio}. Since AudioSet is the largest audio dataset by far, the ablation study on this dataset can best reflect the true performance of DiffRes. The ablation study is performed on mel-spectrograms with 10 ms hop size and a 75\% dimension reduction rate, namely 25 FPS. We perform ablation studies with the following settings: 

% \begin{enumerate}
%     \item \textit{Proposed}: The full proposed DiffRes-based classification system.
%     \item \textit{\textbf{w/o} resolution encoding}: Remove the resolution encoding mentioned in Section~\ref{sec:Monotonic-Frame-warping}.
%     \item \textit{\textbf{w/o} guide loss}: Remove the guide loss~(Equation~\ref{eq:guide-loss}).
%     \item \textit{\textbf{w/o} max aggregation}: Remove the max aggregation function mentioned in the first paragraph of Section~\ref{sec:experiment}.
%     \item \textit{\textbf{w/o} algorithm 1}: Does not apply Algorithm~\ref{algorithm:weight-to-one} on the warp matrix.
% \end{enumerate}

Table~2 in the main paper shows DiffRes does not improve the model performance on $1$~ms setting on most datasets. This may be due to the insufficient receptive field of the convolutions in DiffRes, which is around $41$ time steps. By comparison, the temporal dimension of $\mX$ on AudioSet is $t=3333$ and $t=10000$ with $3$~ms and $1$~ms hop size, respectively. DiffRes may not effectively capture the useful information with only $41$ temporal receptive field in this case. Future work will address this problem by designing the resolution prediction model with a large receptive field. 
% Also, although we use the same $\lambda$ setting in the guide loss~(see Equation~\ref{eq:guide-loss}) in all the experiments, our experiment shows a proper tuning of $\lambda$ can still affect the performance, which might require manual parameter tuning. 

% To sum up, by removing either part in DiffRes, the mAP performance degrade to some extent, as shown in right hand side figure.

% \subsection{Dataset and experiment details}
\begin{table*}[htbp]
\small
\centering
\begin{tabular}{cccccc}
\toprule[\heavyrulewidth]
Dataset & Learning rate & Epoch & Batchsize & \begin{tabular}[c]{@{}c@{}}Learning rate  scheduler \\ (start epoch, gamma, every n epoch)\end{tabular} & GPU(s) \\
\midrule
Audioset       & $1.0\times10^{-4}$ & $30$ & $22 $ & $(11, 0.5, 5) $ & $4$ \\
FSD50K         & $5.0\times10^{-4}$ & $40$ & $15 $ & $(21, 0.5, 5) $ & $1$ \\
ESC50          & $2.5\times10^{-4}$ & $80$ & $32 $ & $(41, 0.95, 1)$ & $1$ \\
SpeechCommands & $2.5\times10^{-4}$ & $60$ & $128$ & $(25, 0.9, 1) $ & $1$ \\
NSynth         & $1.0\times10^{-4}$ & $30$ & $48 $ & $(11, 0.85, 1)$ & $1$ \\
\midrule[\heavyrulewidth]
\end{tabular}
\caption{Hyper-parameter setting. We run all the experiments with an ADAM optimizer~\citep{kingma2014adam} and GeForce RTX 2080 Ti GPU(s).}

\label{tab:hyper-parameters}
\end{table*}

\subsection{Examples}

Illustrated in Figure 4 and Figure 5 of the supplementary materials, this section offers visualizations that depict the impact of DiffRes. These visualizations encompass the fixed temporal resolution mel-spectrogram, frame importance scores, wrap matrix, adaptive temporal resolution spectrogram, and resolution encoding. Such visual aids serve to enhance the comprehension of how DiffRes operates within the context of the audio tagging task.

\label{app-examples}
\begin{figure*}[t]
\centering
  \includegraphics[page=1,width=0.8\linewidth]{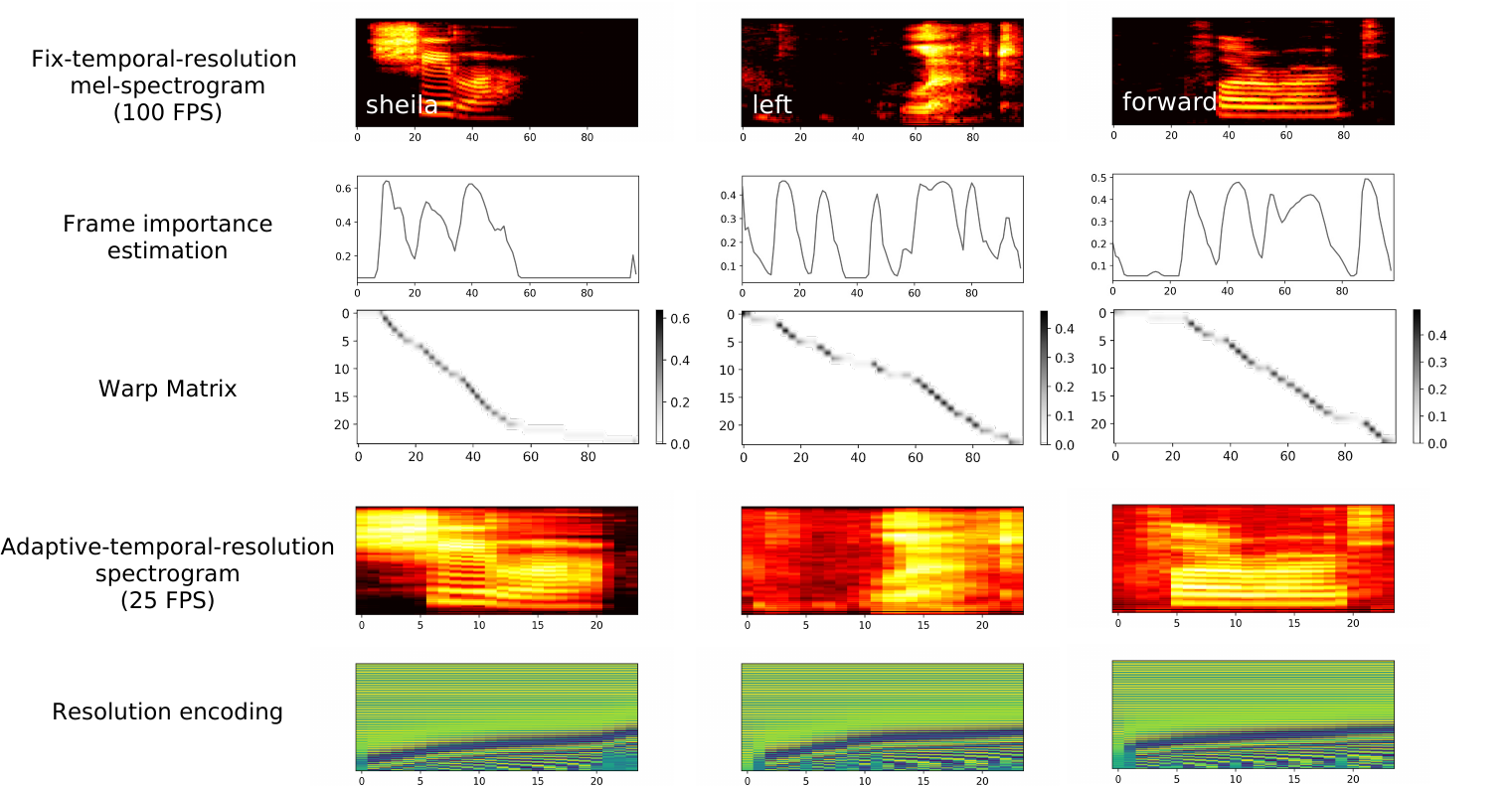}
    \caption{Examples of DiffRes adaptive-temporal-resolution spectrogram on the SpeechCommands dataset.}
    \label{fig-example-sc}
\end{figure*}

\begin{figure*}[t]
    \centering
  \includegraphics[page=1,width=0.8\linewidth]{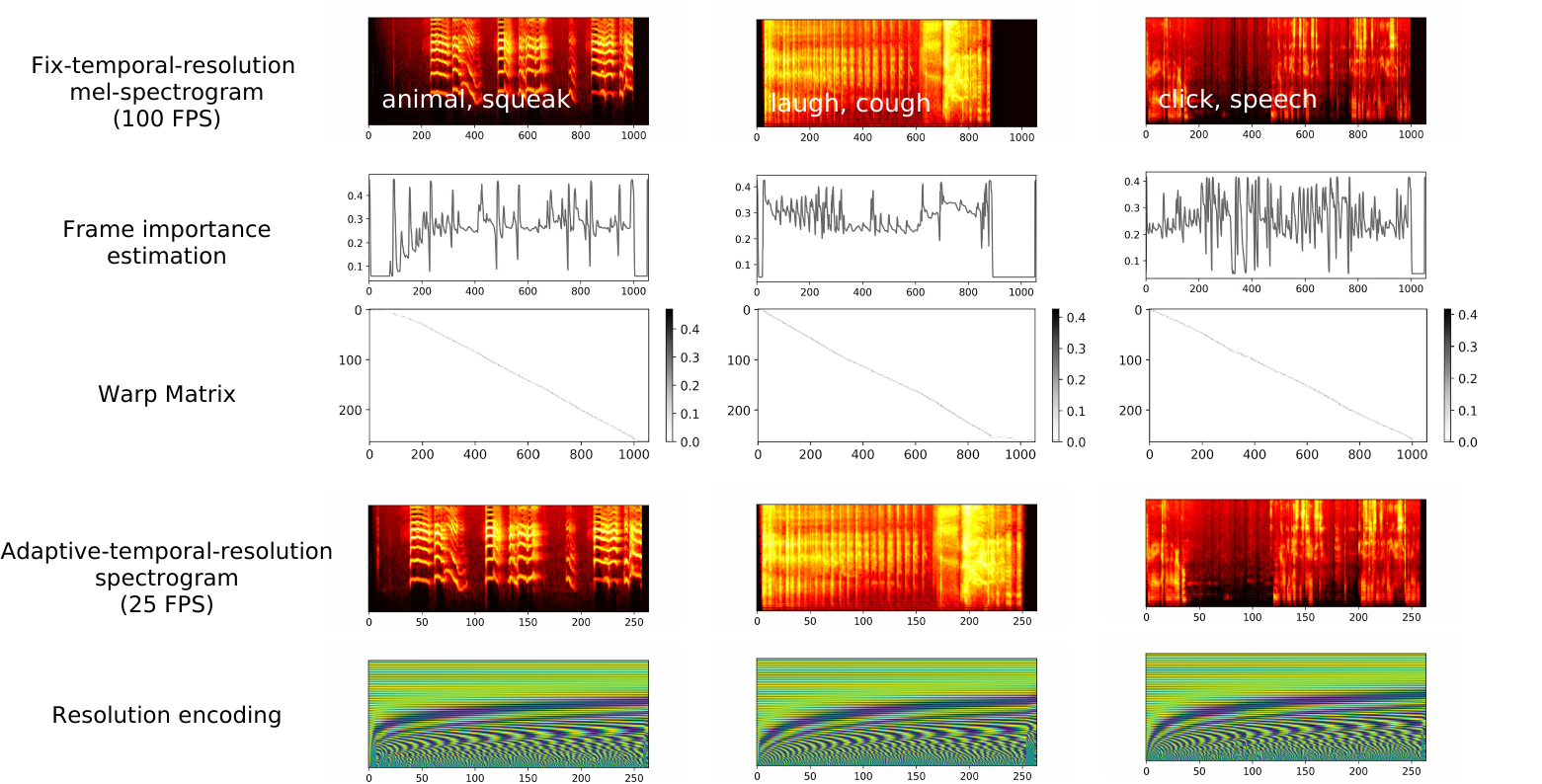}
    \caption{Examples of DiffRes adaptive-temporal-resolution spectrogram on the AudioSet dataset.}
    \label{fig-example-audioset}
\end{figure*}

\clearpage

\bibliography{aaai24}